\documentclass[Physsubmission, Phys]{SciPost}
\hypersetup{
    colorlinks,
    linkcolor={red!50!black},
    citecolor={blue!50!black},
    urlcolor={blue!80!black}
}

\usepackage[bitstream-charter]{mathdesign}
\DeclareSymbolFont{usualmathcal}{OMS}{cmsy}{m}{n}
\DeclareSymbolFontAlphabet{\mathcal}{usualmathcal}

\begin{document}

\begin{center}{\Large \textbf{
JAM-small $x$ helicity phenomenology
}}\end{center}

% TODO: write the author list here. Use initials + surname format.
% Separate subsequent authors by a comma, omit comma at the end of the list.
% Mark the corresponding author with a superscript *.
\begin{center}
Daniel Adamiak\textsuperscript{1$\star$}
\end{center}

\begin{center}
{\bf 1} Department of Physics, The Ohio State University, Columbus, Ohio 43210, USA
* adamiak.5@osu.edu
\end{center}

\begin{center}
\today
\end{center}

% For convenience during refereeing (optional),
% you can turn on line numbers by uncommenting the next line:
%\linenumbers

\definecolor{palegray}{gray}{0.95}
\begin{center}
\colorbox{palegray}{
  \begin{tabular}{rr}
  \begin{minipage}{0.1\textwidth}
    \includegraphics[width=22mm]{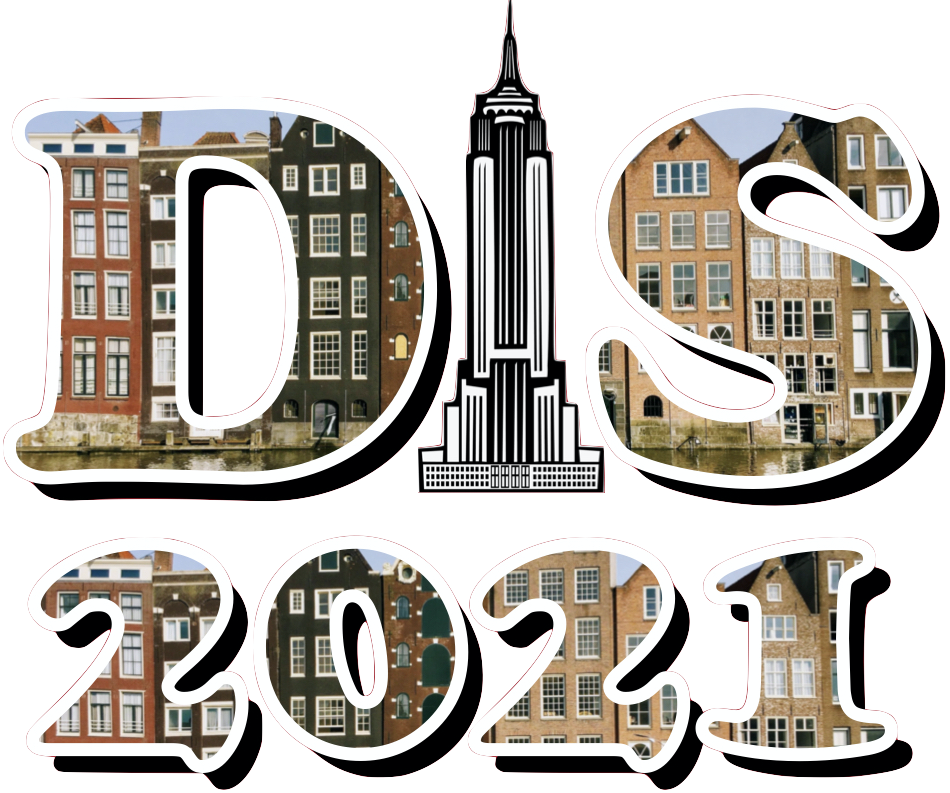}
  \end{minipage}
  &
  \begin{minipage}{0.75\textwidth}
    \begin{center}
    {\it Proceedings for the XXVIII International Workshop\\ on Deep-Inelastic Scattering and
Related Subjects,}\\
    {\it Stony Brook University, New York, USA, 12-16 April 2021} \\
    \doi{10.21468/SciPostPhysProc.?}\\
    \end{center}
  \end{minipage}
\end{tabular}
}
\end{center}

\section*{Abstract}
{\bf
% TODO: write your abstract here.
%The abstract is in boldface, and should fit in 8 lines.
%It should be written in a clear and accessible style, emphasizing the context, the problem(s) studied, the methods used, the results obtained, the conclusions reached, and the outlook. You can add a table contents, recommended if your paper is more than 6 pages long.
 We present the first-ever description the world data on the $g_1^{p,n}$ structure function at small Bjorken $x$ using evolution equations in $x$ derived from first principles QCD. Using a Monte-Carlo analysis within the JAM global framework allows us to fit all existing polarized DIS data below $x<0.1$ as well as predict future measurements of small $x$ $g_1^{p,n}$ at the EIC. This is a necessary step in determining the quark helicity PDFs and, ultimately, the quark contribution to the proton spin.  
}

%\vspace{10pt}
%\noindent\rule{\textwidth}{1pt}
%\tableofcontents\thispagestyle{fancy}
%\noindent\rule{\textwidth}{1pt}
%\vspace{10pt}

\section{Introduction}
\label{sec:intro}
% TODO: write your article here.
This proceedings are based on \cite{adamiak2021analysis}. The proton spin puzzle is one of the largest outstanding facets of QCD, asking how the spin of the proton is decomposed into the angular momentum of its constituents.  The Jaffe-Manohar spin sum rule \cite{Jaffe:1989jz} tell us that the leading contribution to the proton spin is given by
\begin{eqnarray}
	\frac{1}{2} = S_q + L_q + S_G + L_G,
\end{eqnarray}
where $\frac{1}{2}$ is the spin of the proton in natural units, $S_{q(G)}$ is the spin of the quarks (gluons) and $L_{q(G)}$ is the orbital angular momentum of the quarks (gluons). The precise values of these contributions and their functional dependence on the resolution scale, $Q^2$, are still to be determined.
\\
\\
In this work we focus on $S_q$, the spin of the quarks. It can be expressed in terms of the helicity parton distribution functions (hPDFs), $\Delta q$, through
\begin{eqnarray}
	S_q(Q^2) = \sum\limits_q \int\limits_{0}^1 dx \Delta q^+(x,Q^2),
\end{eqnarray}
where $\Delta q^+ \equiv \Delta q + \Delta \bar{q}$ is the sum of the hPDF of a quark and its anti-quark, the sum goes over the contributing quark flavours $u,d,s,$ $Q^2$ is the resolution scale and $x$ is the partonic momentum fraction, which is the same as Bjorken-$x$ at the order we calculate.
\\
\\
While measurements of $\Delta q^+$ are possible down to finitely small $x$, $S_q$ can never be measured directly as it involves probing $\Delta q^+$ at $x=0$, which requires experiments with infinite center-of-mass energies. What we need then is theory that we can trust to evolve in $x$ beyond existing measurements down to $x=0$. These evolution equations, known as KPS evolution, were developed in \cite{Kovchegov:2015pbl, Kovchegov:2016zex, Kovchegov:2016weo, Kovchegov:2017jxc, Kovchegov:2017lsr, Kovchegov:2018znm, Cougoulic:2019aja} (see \cite{Bartels:1995iu, Bartels_1996, Ermolaev:1995fx} for earlier work). This work will describe the process of using KPS evolution to describe the existing data for the $g_1$ structure function, an observable that can be expressed in terms of hPDFs, and make predictions for future EIC measurements of $g_1$.  

\section{Formalism}\label{sec:formalism}
\begin{figure}[h!]
\centering
\includegraphics[width=0.42 \textwidth]{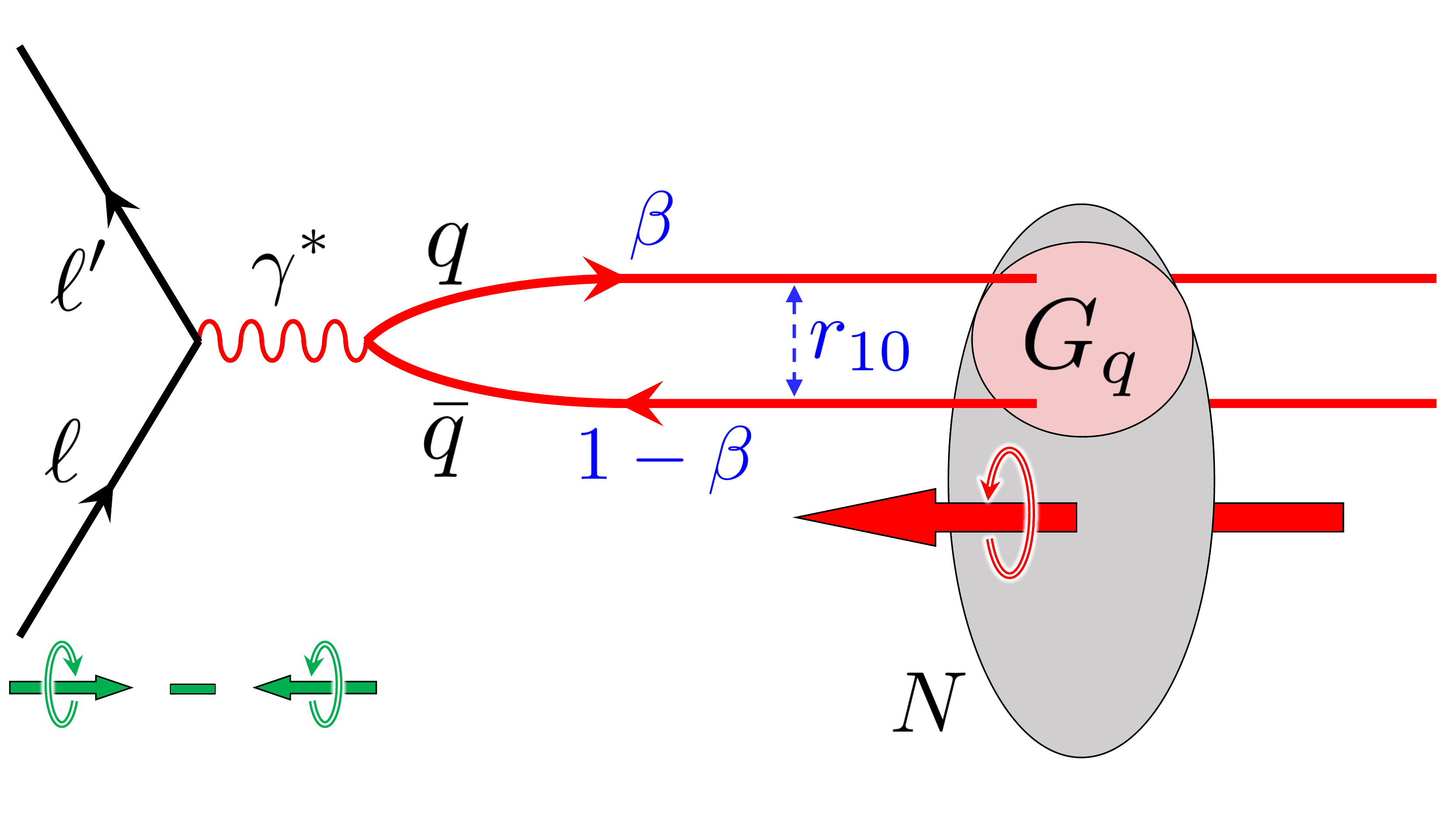}
\caption{Illustration of polarized DIS at small $x$.  The exchanged virtual photon fluctuates into a $q \bar{q}$ dipole of transverse size $r_{10}$, with $\beta$ the fractional energy carried by the less energetic parton in the dipole.  The spin-dependent scattering amplitude of the dipole on the polarized nucleon $N$ is described by $G_q(r_{10}^2, \beta s)$, producing an asymmetry between the cross sections for positive and negative helicity leptons. \cite{adamiak2021analysis}}
\label{fig:DIS}
\end{figure}

At small $x$ and leading order in $\alpha_s$, deep inelastic scattering (DIS) processes are dominated by $q\bar{q}$ pairs piercing the proton medium. Computing these cross-sections is a matter of computing the small-$x$ evolution of the dipole. Similarly, for polarized DIS, where we care about the helicity dependence of our in and out states, processes are once again dominated by quark dipoles piercing the proton medium. However, these dipoles may now exchange helicity information with the proton \cite{Kovchegov:2016zex}, as illustrated in figure \ref{fig:DIS}. This helicity dependent dipole interaction that enters into computations is described by the polarized dipole amplitude, $G_q$, where $q$ runs over the quark flavours. 

The key insight needed to compute the hPDFs is that they may be expressed in terms of the polarized dipole amplitude, $G_q$, through
\begin{align}\label{Dq+3}
    \Delta q^+(x,Q^2) = \frac{1}{\alpha_s \pi^2}
    \int\limits_0^{\eta_{\rm max}} d\eta 
    \int\limits_{s_{10}^{\rm min}}^\eta ds_{10}\, G_q\big(s_{10}, \eta\big),
\end{align}
where the limits on the $\eta$ and $s_{10}$ integrations are given by
    $\eta_{\rm max} = \sqrt{\alpha_s N_c/2\pi}\, \ln(Q^2/x\Lambda^2)$, $\Lambda$ is an infrared cutoff 
and
    $s_{10}^{\rm min} = \max \!\left\{ \eta - \sqrt{\alpha_s N_c/2\pi} \ln(1/x), 0 \right\}$,
respectively. We treat the strong coupling as fixed at $\alpha_s = 0.3$, a typical value for the $Q^2$ range we study.
\\
\\
At leading order in $\alpha_s$, the small-$x$ the evolution of the polarized dipole amplitude is given by the resummation of $\alpha_s \ln^2 \frac{1}{x}$. This is known as the double logarithmic approximation (DLA). Of note, the resummation parameter in unpolarized DIS is $\alpha_s \ln \frac{1}{x}$ and for small-$x$ evolution to apply it is found that $x<0.01$. In order to have the same size parameter in helicity dependent scattering, $\ln^2 \frac{1}{x_{pol}} \approx \ln \frac{1}{x_{unpol}}$, we find that we only need our $x_{pol} <0.1$, allowing us to describe more of the available data.
\\
\\
The evolution of the polarized dipole amplitude closes in the large $N_c$ limit and is given by the following coupled differential equations \cite{adamiak2021analysis}
\begin{subequations} \label{e:HelEv3_f}
\begin{align}
    &G_q (s_{10}, \eta) = G_q^{(0)}(s_{10}, \eta)  
    + 
    \int\limits_{s_{10}+y_0}^{\eta} d\eta' 
    \int\limits_{s_{10}}^{\eta' - y_0} d s_{21} \: 
    \big[ \Gamma_q (s_{10} , s_{21} , \eta') + 3\, G_q (s_{21}, \eta') \big] , \nonumber\\\\
    &\Gamma_q (s_{10} , s_{21} , \eta') = G_q^{(0)}(s_{10} , \eta')   
    + \int\limits_{s_{10}+y_0}^{\eta'} d\eta''
    \int\limits_{s_{32}^{\rm min}}^{\eta'' - y_0} ds_{32} \, 
    \big[ \Gamma_q (s_{10} , s_{32} , \eta'') + 3\, G_q (s_{32}, \eta'') \big] . \nonumber\\
\end{align}
\end{subequations}
$\Gamma_q$ is an auxiliary function that obeys it's own evolution equation that mixes with $G_q$. Importantly, this system is closed and can be calculated numerically. $y_0:=\sqrt{\alpha_s N_c/2\pi} \ln \frac{1}{x_0}$ ensures that evolution only begins below a sufficiently small $x<x_0$. $G^{(0)}_q$ is flavour dependent initial condition that follows the Born-inspired form:
\begin{eqnarray}\label{eq_IC}
	G^{(0)}_q(s_{10},\eta) = a_q \eta + b_q s_{10} + c_q,
\end{eqnarray} 
where $a_q, b_q$ and $c_q$ are flavour dependent parameters that need to be fit to data. 
\\
\\
With the formalism in place, the only missing piece is deciding how to constrain the initial condition, $G^{(0)}_q$. The total spin, $S_q$, should be dominated by the three light quark hPDFS, so we need to at least be able to determine $\Delta u^+, \Delta d^+$ and $\Delta s^+$ separately. We therefore need to fit to data of at least three observables, expressible as linearly independent combinations of the hPDFs to nail each down separately. 

\section{Observables}
Polarized DIS gives us access to three prime candidates for determining the hPDFs; the structure functions $g_1^p, g_1^n$ and $g_1^{\gamma z}$. These are expressable as linearly independent functions of the hPDFs and can be extracted from the data with minimal bootstrapping of additional theories, i.e. there is no need to invoke fragmentation functions or similar structures that would have to be fit simultaneously with $G_q^{(0)}$. Unfortunately, there is currently no data for parity violating DIS and thus no data for $g_1^{\gamma Z}$.
\\
\\
Never-the-less, we will extract as much information as we can from the proton and neutron $g_1$ structure functions to demonstrate that our formalism can describe existing data, as well as make meaningful predictions about these structure functions.
\\
\\
We can then generate pseudo-data to demonstrate the impact the electron-ion collider (EIC) will have on our predictions. Finally, we will impose an artificial third constraint on the hPDFs that will demonstrate how the separate hPDFs can be in principle be extracted once $g_1^{\gamma Z}$ is measured at teh EIC.
\\
\\
The proton $g_1^p$ structure function is given by
\begin{eqnarray}
	g_1^p(x,Q^2) = \frac{1}{2} \sum\limits_q e_q^2 \Delta q^+(x,Q^2) 
\end{eqnarray}
Even if we cannot access each distribution individually at the moment, we may re-purpose the undetermined constants in \eqref{eq_IC} to be constants for the $g_1$ structure function, writing 
\begin{eqnarray}\label{eq_g1_IC}
	g_1^{(0)} = a_{g_1} \eta + b_{g_1} s_{10} + c_{g_1}.
\end{eqnarray} 
This is possible because $g_1^{p,n}$ is a linear combination $\Delta q^+$ and thus follows the same evolution equation. In other words, we can try to describe $g_1^{p,n}$ with only three constants each, instead of the nine needed to describe the hPDFs. 
\\
\\
The $g_1^{p/n}$ structure functions may then be extracted directly from double spin asymmetries in polarized DIS. At large $Q^2$ these asymmetries are simply related to the structure functions through $A_{||}\propto A_1 \propto g_1/F_1$. The structure function $F_1$ is taken from the JAM global analysis \cite{Cocuzza21,Sato:2019yez}.
\\
\\
Now that we know which data to look at, we may perform a fit of the small-$x$ formalism by employing the JAM framework: Monte-Carlo generation of fit parameters that tend towards a minimum $\chi^2$ through Bayesian updates \cite{Sato:2019yez, Moffat:2021dji}.

\section{Results}
\begin{figure}[h!]
\centering
\includegraphics[width=0.483\textwidth]{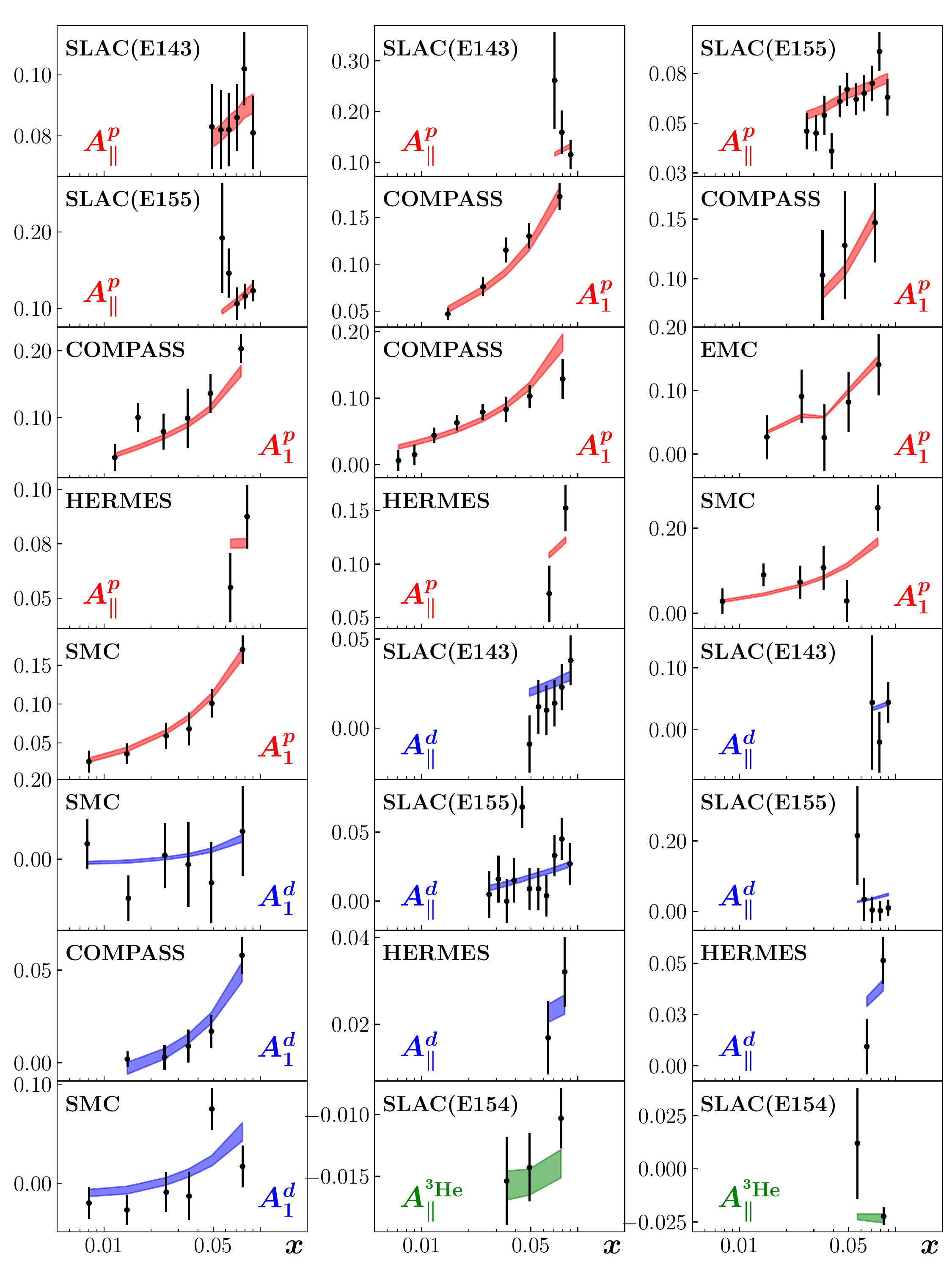}
\caption{Comparison of longitudinal double-spin asymmetry data (black) to our fit on proton (red), deuteron (blue), and $^3$He (green) targets at $x \! < \! 0.1$ and $Q^2 \!\! \in \! [1.73, 19.70]$~GeV$^2$ \!\! with the JAMsmallx fit.}
\label{f:A1}
\end{figure}

\begin{figure}[h!]
\centering
\includegraphics[width=0.83\textwidth]{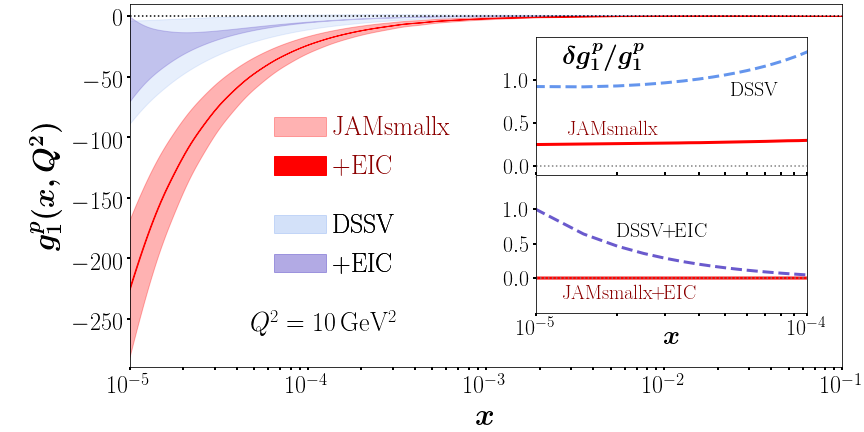}
\caption{
\mbox{JAMsmallx} result for the $g_1^p$ structure function obtained from existing polarized DIS data (light red band) as well as with EIC pseudodata (dark red band). For comparison, we include $g_1^p$ from the DSSV fit to existing data~\cite{deFlorian:2014yva, deFlorian:2019zkl} (light blue band) and with EIC pseudodata at $\sqrt{S}=45$ and 141~GeV~\cite{Aschenauer:2020pdk} (light purple band). The inset gives the relative uncertainty $\delta g_1^p/g_1^p$ for each fit at small $x$.}
\label{f:g1}
\end{figure}

\begin{figure}[h!]
\centering
 \includegraphics[width=0.425\textwidth]{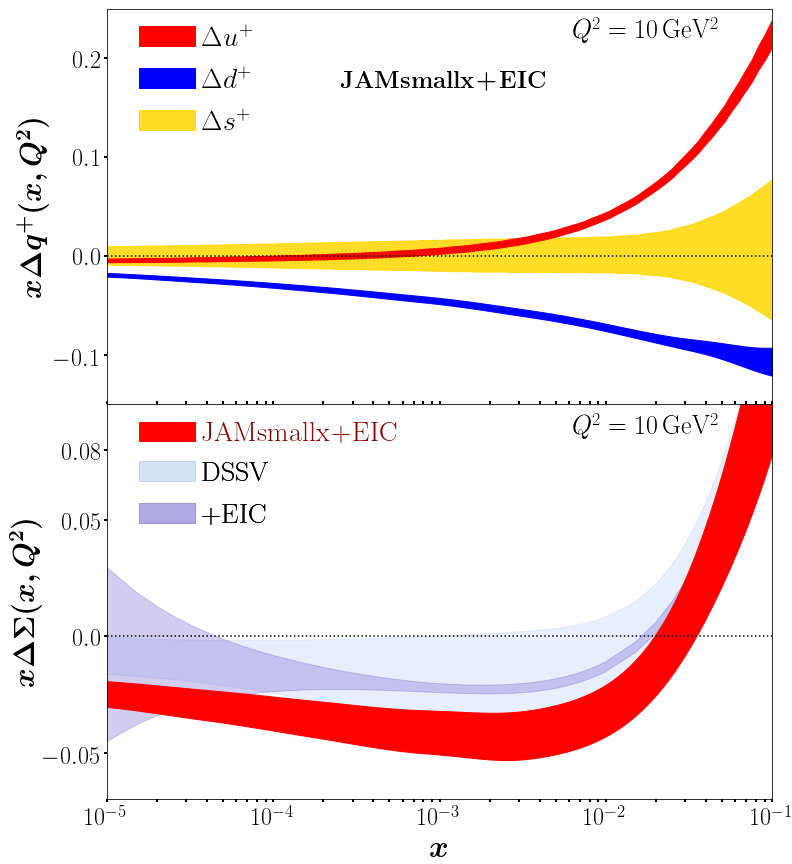}
\caption{{\bf (Top)} Fitted helicity PDFs $x\Delta q^+(x,Q^2)$ from the current JAMsmallx fit to existing polarized DIS data and EIC pseudodata for $A_{||}$ and $A_{\rm PV}$ at $x<0.1$. {\bf (Bottom)} The result for $x\Delta\Sigma(x,Q^2)$ from the same fit (red), compared with that from the DSSV analysis with~\cite{Aschenauer:2020pdk} (light purple) and without~\cite{deFlorian:2014yva, deFlorian:2019zkl} (light blue) the EIC pseudodata.}
\label{f:hPDFs_Sigma}
\end{figure}

There are constraints that limit the data we may try to describe. We have already discussed that the largest $x$ at which this formalism applies is $x=0.1$. We also restrict the data we analyse to $Q^2>m_c^2\approx1.69 GeV^2$ and $s>4GeV^2$. The data sets included are from the SLAC~\cite{Anthony:1996mw, Abe:1997cx, Abe:1998wq, Anthony:1999rm, Anthony:2000fn}, EMC~\cite{Ashman:1989ig}, SMC~\cite{Adeva:1998vv, Adeva:1999pa}, COMPASS~\cite{Alexakhin:2006oza, Alekseev:2010hc, Adolph:2015saz}, and HERMES~\cite{Ackerstaff:1997ws, Airapetian:2007mh} experiments. 
\\
\\
Using the initial conditions \eqref{eq_g1_IC} in the evolution equations \eqref{e:HelEv3_f} to calculate the $g_1$ structure function gives us the description of the data shown in Fig. \ref{f:A1}. The $\chi^2/Npts$ for these fits is 1.01, demonstrating that this formalism can successfully describe existing data.
\\
\\
We are now able to predict the behaviour of $g_1^p$ down to $x=10^{-5}$ (shown by the light red band in Fig. \ref{f:g1}). Contrast this with DGLAP descriptions, e.g. DSSV (light blue band), that parameterize the $x$ behaviour. The consequence of this distinction can be seen in the inset plots showing $\delta g_1^p/g_1^p$, where we demonstrate that we maintain good control over the relative uncertainty of our prediction, well beyond the value of $x$ where there will be EIC data. The dark red and light purple also show the impace of EIC pseudo-data on our small-$x$ prediction and DSSV respectively. The EIC pseudo data is generated from the $g_1^p$ fit. 
\\
\\
Lastly, we present a preliminary extraction of the hPDFs and and use them to calculate the quark spin contribution $\Delta \Sigma (x,Q^2) = \sum\limits_q \Delta q^+$. In order to be able to extract the hPDFs we impose an artificial constraint that $\Delta s^+ =0$ so that we may generate pseudo-data for parity violating DIS under this zero-strangeness assumption. The extraction of the hPDFs and $\Delta \Sigma$ are given in \ref{f:hPDFs_Sigma}.

\section{Conclusion}
In order to solve the proton spin puzzle, it is necessary to describe the hPDFs down to zero $x$. We have shown that KPS evolution presents great progress on this front. Not only may it be used to describe existing double spin asymmetries (Fig. \ref{f:A1}), but the uncertainty also remains under good control as we extrapolate to smaller $x$ (Fig. \ref{f:g1}).
\\
\\
Existing inclusive DIS does not completely constrain all the hPDFs, but we are still able to demonstrate the capability of the JAM smsall-$x$ framework to extract them by generating pseudo-data with artificial constraints. In future work, we will substitute these artificial constraints by physical ones when exploring observables that can be measured in semi inclusive DIS.
\\
\\
None-the-less, we managed to describe $g_1^p$ and make predictions that could be measured at the EIC (Fig. \ref{f:g1}). Moreover, we show the impact that the EIC would have on constraining our description of the structure functions and the hPDFs and on extrapolation of our results down to even smaller values of x.

\section*{Acknowledgements}

These proceedings are based on the work in ref \cite{adamiak2021analysis}. This work has been supported by the U.S. Department of Energy, Office of Science, Office of Nuclear Physics under Award Number DE-SC0004286.

\nolinenumbers

\end{document}